# From Prompt Engineering to Prompt Craft

**Joseph Lindley**
ImaginationLancaster, School of Design
Lancaster University, UK
j.lindley@lancaster.ac.uk

**Roger Whitham**
ImaginationLancaster, School of Design
Lancaster University, UK
r.whitham@lancaster.ac.uk

**ABSTRACT**

This pictorial presents an ongoing research programme comprising three practice-based Design Research projects conducted through 2024, exploring the affordances of diffusion-based AI image generation systems, specifically Stable Diffusion. The research employs tangible and embodied interactions to investigate emerging qualitative aspects of generative AI, including uncertainty and materiality. Our approach leverages the flexibility and adaptability of Design Research to navigate the rapidly evolving field of generative AI. The pictorial proposes the notion of *prompt craft* as a productive reframing of prompt engineering. This is comprised of two contributions: (1) reflections on the notion of materiality for diffusion-based generative AI and a proposed method for a craft-like navigation of the latent space within generative AI models and (2) discussing interaction design strategies for designing user interfaces informed by these affordances. The outcomes are presented as strong concepts or intermediate knowledge, applicable to various situations and domains.

**INTRODUCTION**

The rapid evolution of diffusion-based AI image generation systems presents unique challenges and opportunities for design research. This pictorial explores these affordances through three practice-based Design Research projects conducted in 2024, focusing on various versions of Stable Diffusion. Our approach, synonymous with Research through Design (RtD), emphasises tangible and embodied interactions as key elements of inquiry.

Practice-based Design Research offers a promising method for understanding emerging qualitative aspects of generative AI, including uncertainty [1] and materiality [6]. The inherent flexibility and reflexivity of design-led research are particularly valuable in this fast-moving field [2], where the state-of-the-art can shift during the course of an investigation. Moreover, the metaphorical material properties of AI systems [4,9] underscore the relevance of interpretive approaches over positivist-leaning controlled experiments.

Gaver, Krogh, Boucher, and Chatting aptly describe the unpredictable nature of such research with the Yiddish proverb "Der mentsh trakht und got lakht" (People plan and God laughs) [5]. They propose strategies for navigating this uncertainty, including embracing technical affordances to suggest new directions, appreciating the value of research programmes over projects, and viewing research as a journey rather than a quest. These strategies have both informed and manifested in our research.

While distinct, our three projects share a common thread in developing novel input, output, and interaction mechanisms for diffusion-based generative AI models. These were tailored to specific contexts and technical limitations. As outcomes of our approach, the contributions we proffer are presented as strong concepts or intermediate knowledge [7].

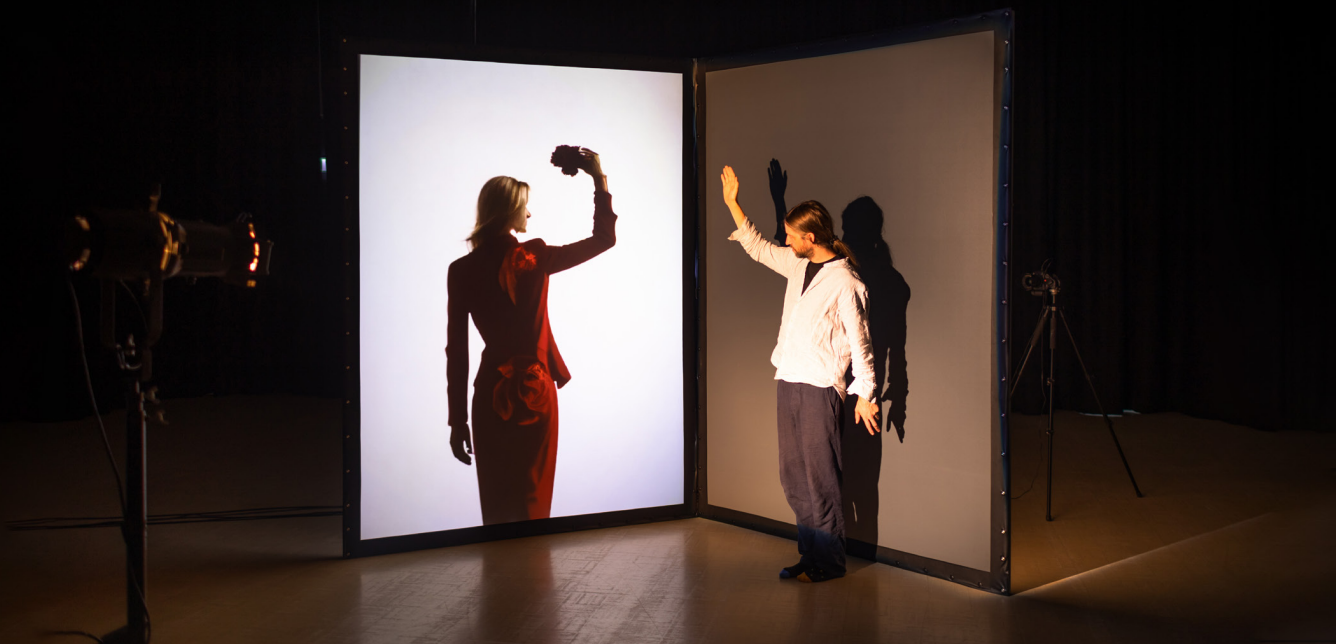

## SHADOWPLAY V1

The first project we present, Shadowplay, intended to create an art exhibit that used participants' bodies as tangible input for an AI image generator. The project built upon the Entoptic Media Camera, a project that used AI image-to-image generation to explore the uncertainty inherent within AI models [1]. When used in image-to-image mode, generative AI systems are passed an input image along with an optional text prompt. In our case we were using a live stream from a camera as the input image, allowing participants to use their bodies as the main content of the input image. Early experiments revealed that the resulting outputs lacked consistency making the aesthetic of the interaction overwhelming, confusing, and unsatisfactory.

Searching for ways to temper this discomfort led us to experiment with using bright lights, allowing participants to cast a shadow on a surface as opposed to their body being captured directly. Capturing the shadows cast by the participants meant that all the inputs were mainly monochrome and were high contrast. Giving the input images these attributes constrained the potential of the AI model's latent space [10] and resulted in a more refined exhibition experience with a more consistent aesthetic. Given our workflow involved using a camera to capture images comprised of light and dark areas, we refer to this approach as *light prompting*.

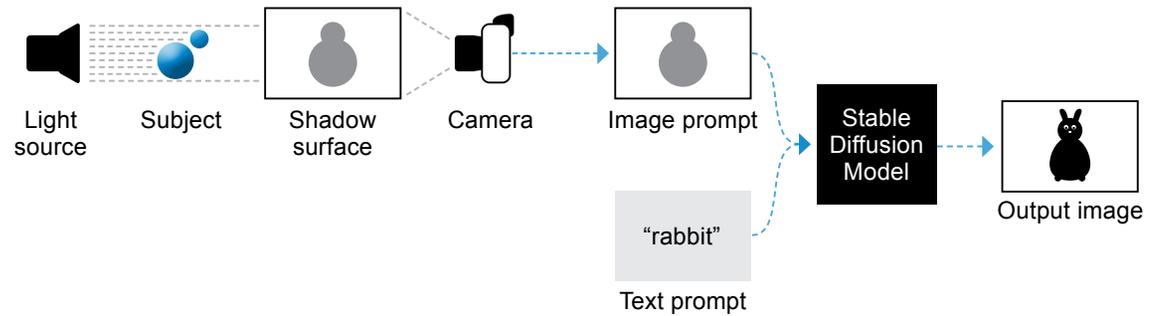

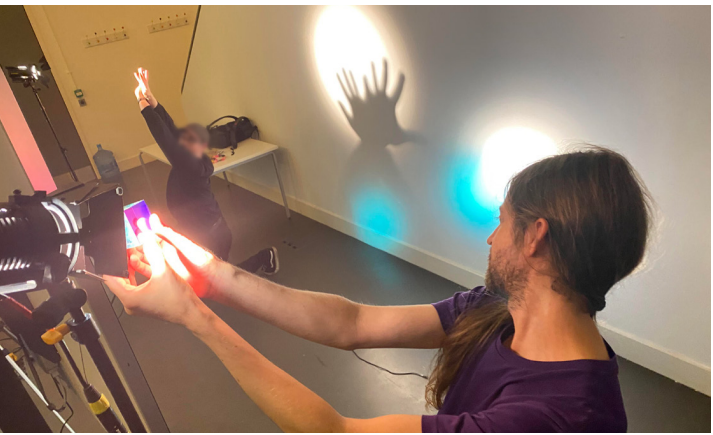

Fig 2: Casting shadows with hands and a prism

## Leveraging Light Prompting

The shaping effect of light prompting on the model's latent space is strong. For example, in early experiments and before we introduced any textual prompts to the work, we saw many images reminiscent of war photography. Our assumption is that many of the high contrast, monochromatic images in the training set depict scenes of war. Hence, when a high-contrast monochromatic image is used as input, the model is likely to recreate war imagery. This was not an aesthetic we wished our exhibit to have, hence using text prompts to steer the model away from this type of imagery became a focus.

To harmonise with the aesthetic of our high-contrast shadow-based images, we began to explore text prompts that referenced shadow puppetry. At this point the aim was to allow participants to use their bodies to cast shadows, the AI would augment and stylise those shadows resulting in AI-facilitated and embodied shadow puppetry. In our experiments the most evocative and aesthetic images arose when we came to include the name Lotte Reiniger—an artist and director who was a prominent pioneer of shadow/silhouette animation—within the text prompt.

Ultimately, we chose to make this prompt central to this version of Shadowplay. Referring to Lotte Reiniger in this way both brought a playful and striking aesthetic but also allowed us to shine a critical light on the complexities that generative AI brings to notions such as authorship, ownership, and intellectual property. Using Reiniger's name to create images that resembled her body of work and exemplifies why creators and artists profess concern about the impact of AI on their profession. While the matter is too complex to explore in this pictorial, we discussed the ethics of this decision coming to the view that no discernible harm would be caused, in part because Reiniger died in 1981.

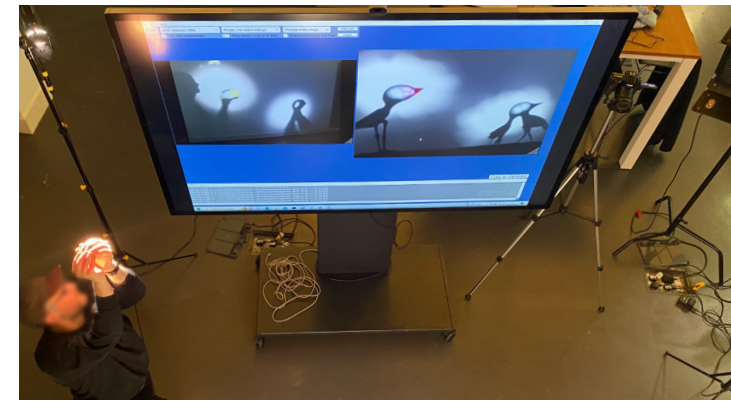

The input and output image side-by-side on a screen

**Exhibition at TEI2024**

Shadowplay's inaugural exhibition was at the 2024 TEI conference in Cork, Ireland. The installation comprised of a light, a camera, a computer running the software, and two screens. The software was built using node-based programming environment TouchDesigner and it leveraged a component TDDiffusionAPI [3] to interface with the AI model (Stable Diffusion 2.1 Turbo). The AI model was served via an instance of Automatic1111 which was hosted on a cloud server.

**Reflecting on the Exhibition**

Interestingly, although instructed to only cast shadows, participants would frequently stand directly in front of the camera. This disrupted our planned use of light prompting but still yielded aesthetic and interesting results. We learned that Shadowplay's ability to engage participants emerged from an evident connection between input and output. While participants knew their actions directly influenced the output, there was also a constant element of unpredictability, often manifesting in striking and grotesque imagery. This fostered a constant sense of anticipation in the audience. Through the course of the exhibition, we began to edit the underlying text prompt. We would incorporate requests ranging from objects like flowers or robots through to more abstract concepts like happiness or anxiety. While we editing the prompt we maintained a reference to Lotte Reiniger to retain a constant aesthetic.

**Trading Intrigue for Stability**

AI image generation models like Stable Diffusion use a 'seed' to generate random noise which drives the model's ability to create infinite variations of images. After experimentation we decided to use a random seed for each frame of the generated imagery. This choice represented a trade-off. If we used a fixed seed the output images were stable, and participants could gradually evolve and shape the imagery and between frames. However, at the frame rate (1 fps) the experience felt somewhat boring. With a random seed, the experience was more immersive and intriguing, although participants had less control, and the output images were less stable.

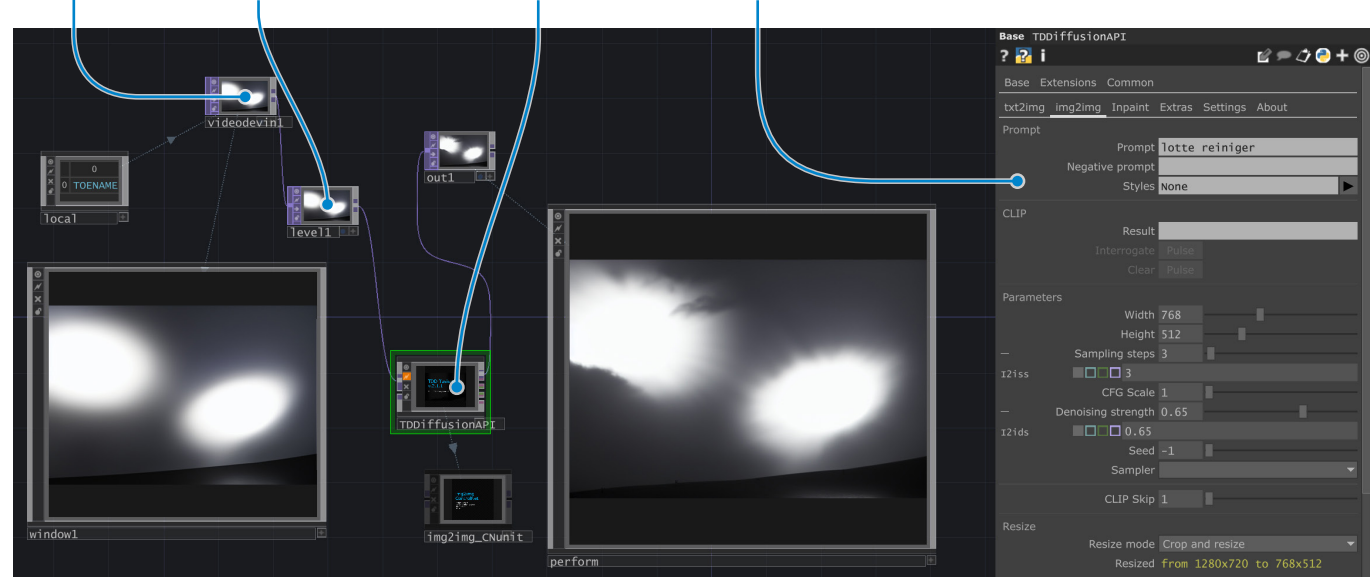

**TouchDesigner network for image processing**
image input, exposure adjustment, a stable diffusion component and settings

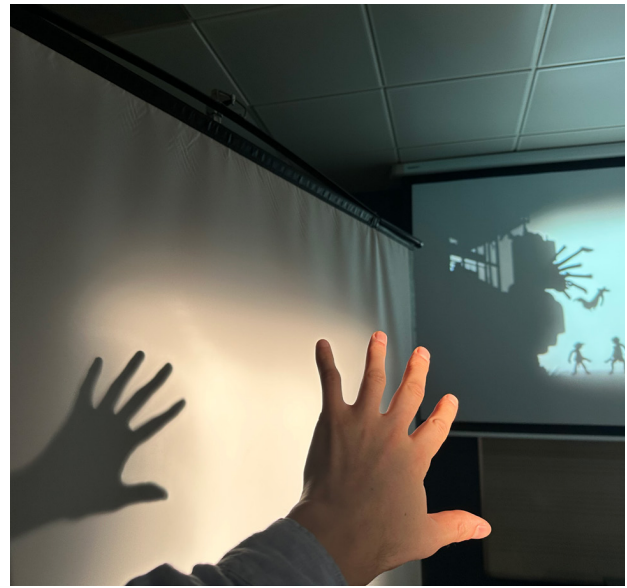

The input and output screens within the installation

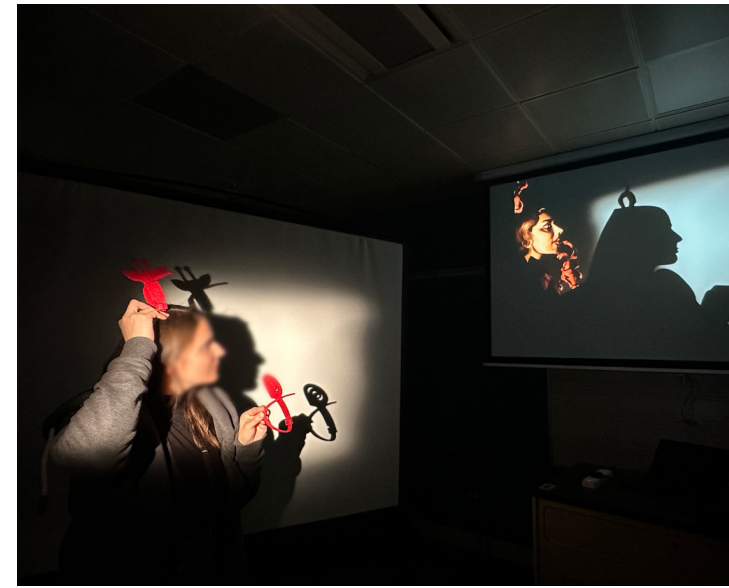

A participant using their body and shadow as input

## CARDSHARK

The second project we present is named Cardshark and formed around the opportunity to use AI in delivering a workshop that was seeking to generate ideas for novel musical instruments. The aim was to use AI to help empower the participants, who were individuals living with Young Onset Dementia (YOD), to express themselves. Loss of language and reduction of ability to articulate ideas are typical symptoms of YOD, so we sought to use AI to temper those symptoms within the context of the workshop. The workshop had 25 participants who were part of an existing YOD support group. We developed Cardshark around the to the purpose of the workshop, the venue, and the anticipated needs of the participants.

Behind the scenes, Cardshark uses traditional text-to-image prompting. However, we developed a tangible and accessible input modality using customised *Prompt Fragment Cards*. These cards were placed into a *Prompt Arena*, and finally plugged into a *Meta Prompt*. Results were generated in near real-time (~1 fps) and displayed immediately for all members of a group. In contrast to Shadowplay, Cardshark used a stable seed, meaning that each image is consistent with the last one generated, allowing for fine-tuning of output images. Our goal was to allow participants to combine fragments that interested them, give a consistent quality to output images, minimise distracting elements, encourage iterative crafting, and enable individual and collaborative making with minimal support.

An initial research visit with the group strongly informed our approach to designing the interactions the Cardshark system afforded. The group knew each other well and thrived on having conversations with one another. Members of the group articulated preferences for experiences that helped them focus their attention and "kept them busy". Most were users of digital technology (such as smartphones), but some struggled to express themselves comfortably using written words.

In this pictorial there is not the scope or space to report on the successes and failures of the workshop for participants. Our focus here is on the process of crafting the interaction between participants and an AI image generation tool.

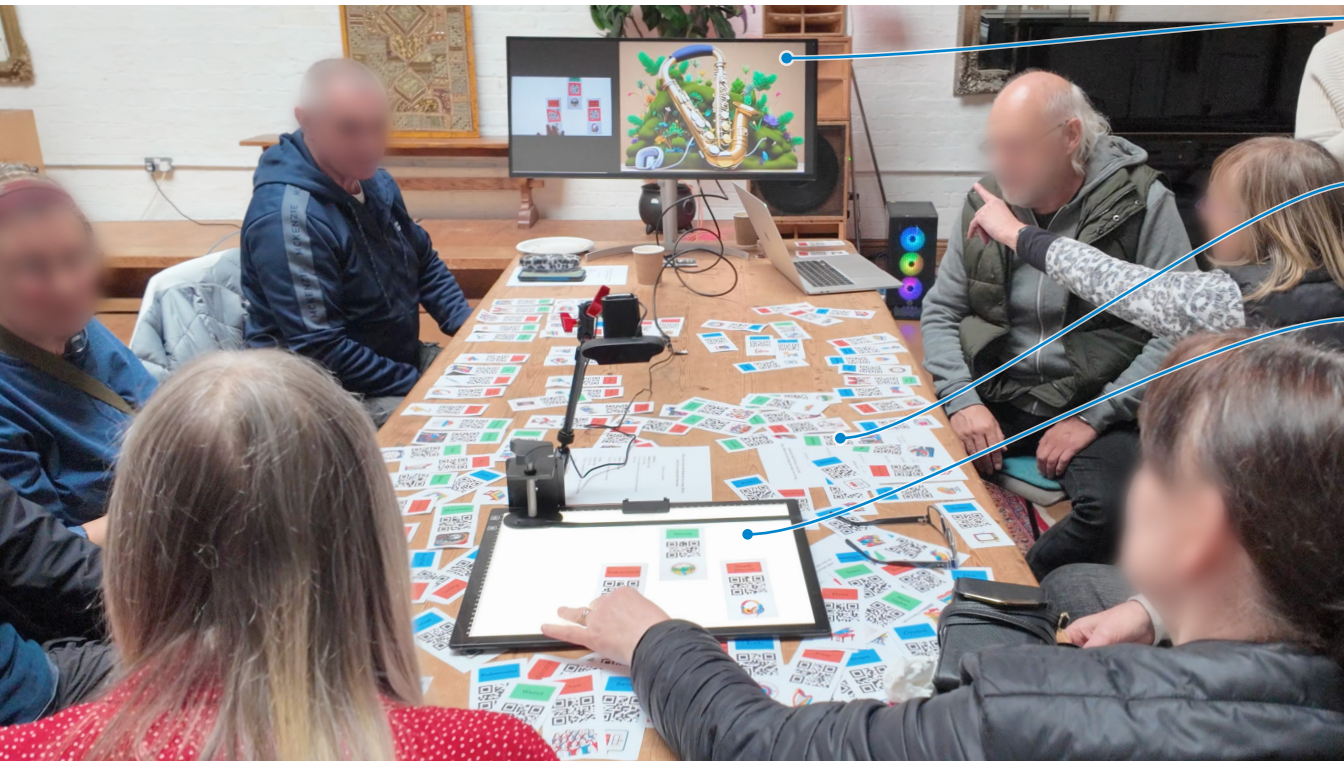

A group using Cardshark at within the YOD Workshop, April 2024

*Main display*
The image currently being crafted by the group is displayed at the head of the table, updating once per second, and as the system uses a stable seed the output image remains constant unless fragments are changed.

*Prompt Fragment Cards*
Each card comprised a picture of an object or concept with a text label, and a QR code that the system used to detect which fragments were in the Prompt Arena.

*Prompt Arena*
An illuminated lightbox with a camera suspended above provided a clearly delineated collaborative interface that could be passed around the table. Placing Prompt Fragments on the Arena immediately updated the resulting output image on the main display. Sliding fragment cards up and down altered their weighting within the prompt.

*Facilitation*
The facilitator had a bespoke interface allowing them to store images created by participants, add titles and notes to them, and print take-away postcards of the generated images. Some participants used the system and understood it without being asked to and as the workshop progressed most participants quickly mastered the system and did not need facilitator support.

**Crafting the *Latent Possibility Space* of Cardshark**

For Cardshark to work we needed outputs to be engaging and playful, but visually consistent and uncluttered. To achieve this we engaged a craft-like process to arrive at a *meta-prompt* that would allow us to access the useful part of the model's latent possibility space.

The blue area represents the intended relative size of the model's latent possibility space

*Model selection* → *Prompt fragment generation* → *Meta-prompt creation* → *Fragment pruning*

The choice of model influenced the overall possibility space, style, speed of generation, and consistency. We tested a range of models with the same prompt to identify a candidate. We used the *Vibrant Horizon Turbo XL* model.

Our fragments comprised concepts relating to musical instruments, including specific instruments (e.g., cello), components (e.g., string) and wildcard terms (e.g., balloon). We batch-tested these with the model to finetune the terms.

We iteratively developed our meta-prompt by testing various linking phrases in combination with our prompt fragments. The process was highly iterative, involving back and forth testing of combinations of fragments and prompts. Emerging from this process, crucial aesthetic, instructional and linking terms emerged. For example, the prompting for a hybrid was more effective than innovative or creative. As candidate meta-prompts emerged, we batch tested these with seeds and fragment combinations. This iterative and craft-like process resulted in an assemblage of model, fragments, and meta-prompts that constrained the latent possibility space of the model while still provisioning the necessary amount of creative possibilities.

As we experimented with the resulting prompt space, we removed some fragments that exhibited distracting effects. Some fragments were also very dominant when used in certain combinations, requiring adjustments in weighting.

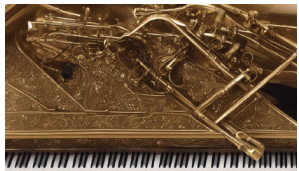
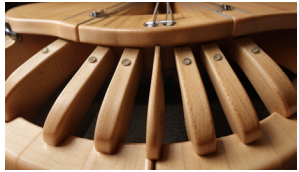
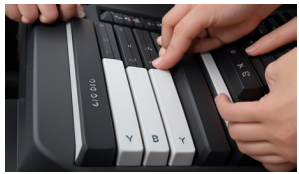

Pilot prompts tested with several different models

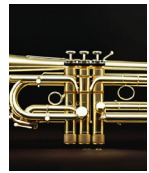 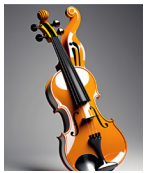
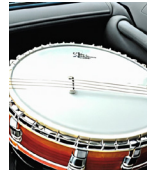 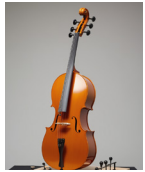
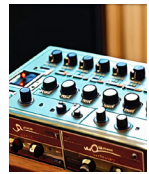 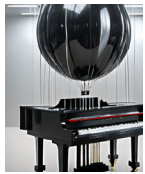

Instrument fragments tested individually without a meta prompt

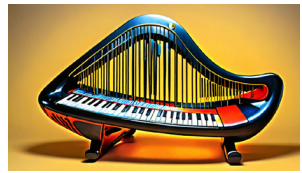
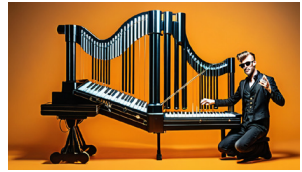
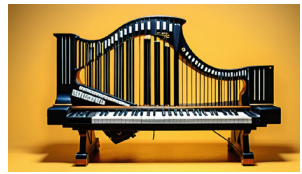

Experiments with negative prompts and other parameters such as classifier free guidance

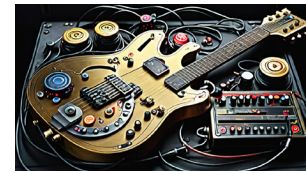
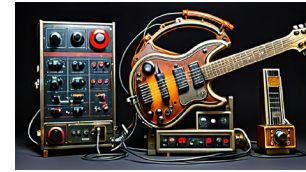
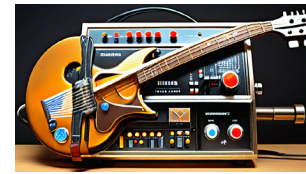

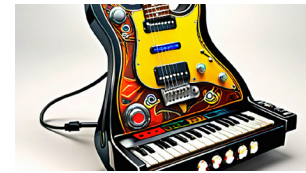
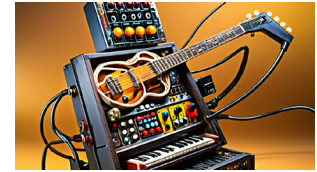
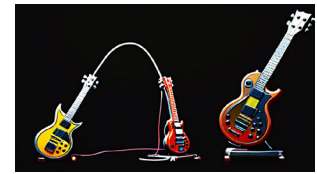

Experiments with different connective terms (e.g., hybrid, fusion, innovative) to prompt combined instruments

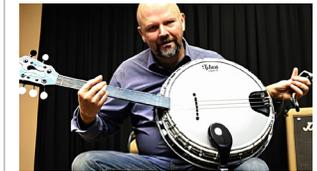
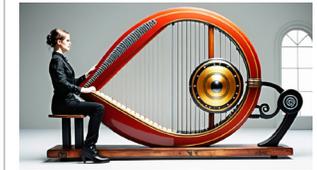
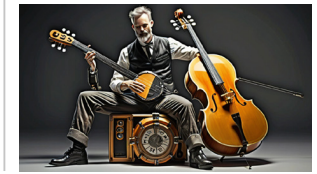

Banjo and harp fragments often brought people with them

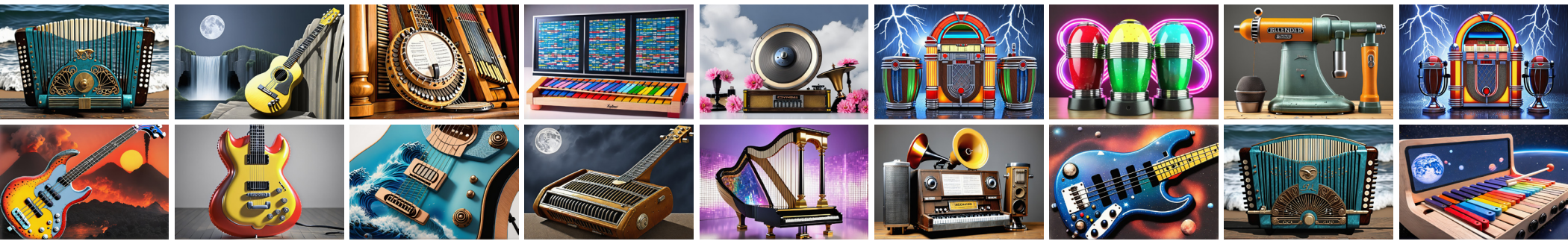

A selection of Cardshark images selected by participants to be saved and printed

**Meta-prompt interaction**

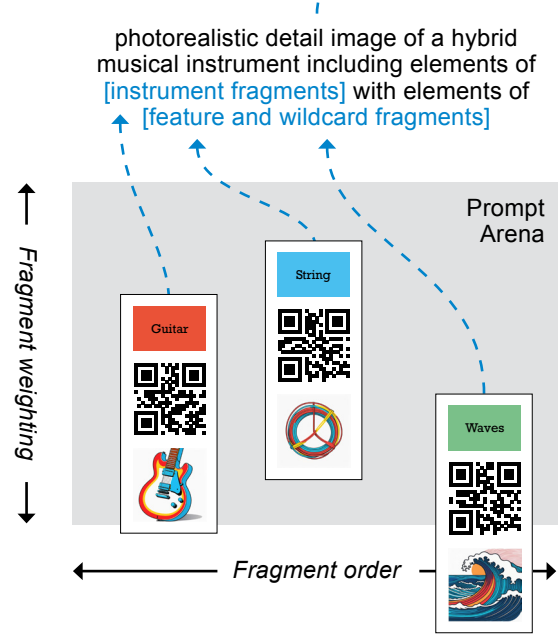

photorealistic detail image of a hybrid musical instrument including elements of [instrument fragments] with elements of [feature and wildcard fragments]

The prompt arena presented as a lightbox, webcam and fragment cards proved successful in practice. Workshop participants quickly understood the idea of adding fragments and adjusting their relative weighting.

**Responsivity Leading to Intuitive Interaction**

Cardshark and the workshop that employed it was well-received by participants. A representative of a dementia charity in attendance commented on the empowerment evident in the session, both in engaging participants in creative making, and enabling them to articulate their ideas and experiences.

While the success or failure of the workshop is not the focus of this pictorial, participants engagement with our novel tangible interaction system demonstrated it was an accessible way to interact with the model's latent possibility space. Over the course of the workshop some participants interacted with the system extensively, and discovered unexpected behaviours of prompt fragments. For example, finding that some fragments had consistently little effect, or needed to receive a high weighting to be useful.

Compared to Shadowplay V1, the Cardshark system's higher frame rate (~1fps) allowed us to design interactions with real-time feedback for users. This responsivity led us to develop the the *prompt arena* and *fragment cards*, which, in turn, empowered our participants to operate the system autonomously and without any facilitator intervention.

**Crafting the Possibility Space**

Cardshark offers a playful, collaborative and accessible interface for creating text prompts for generative AI. To deliver this involved conventional user research techniques, e.g., prototyping and heuristic evaluation informed by early engagement with potential users. In contrast, the co-evolution of model choice, fragments, meta-prompt and possibility space was less conventional. Within the context of rapid innovation and evolving perspectives on interaction design for AI, this deserves reporting in isolation.

We developed a craft-like approach to working with the AI model. This included moments of systematic exploration (e.g., testing fragments in combination and with different weightings) but that could only follow creative moments (e.g., experimenting with different connecting terms and producing fragment lists). The process was iterative, involving fixing some parameters while modulating others, with unknown outcomes. We developed a practice of oscillation between modes of manually developing candidate prompts and then automatically testing them by batch producing images from a matrix of fragments. We realised this became necessary as we discovered the 'soft edged' of the model's latent possibility space. Like a craftsperson exploring the properties of a new material through trial and error, gradually building up understanding through non-linear experimentation.

**SHADOWPLAY V2**

The third project we report on is Shadowplay V2, an iteration on the original concept that implements StreamDiffusion [8], which yields a generation speed of around 12 frames per second (36 times faster than Shadowplay V1). This technical leap leads to a real-time experience, leading us to explore and develop several ways to incorporate new elements of dynamism into the installation. These include real-time embodied control of some of the AI's parameters, dynamically changing text prompts, and reactive sound and music elements.

Shadowplay V2 itself has, so far, had two installations, one at *Electromagnetic Field* (an outdoor maker festival with ~3000 attendees) at *Storytellers + Machines*, a transdisciplinary conference on creative AI. The significant step up in performance afforded by StreamDiffusion led to new features, user interfaces, and ways of interacting with Shadowplay, characterised by dynamism. While in Shadowplay V1 we chose to have a random seed for each frame to make each frame unique, in Shadowplay V2 the seed is static, but many other factors add variety and dynamism to the overall experience of interaction.

**Diffusion Amount**

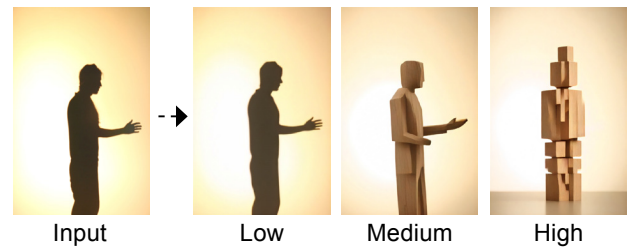

The first of these types of dynamism is the balance between what the camera is looking at and AI in any given frame. We used participants' movements to modulate this. As participants made bigger and quicker movements the subsequent output images would increasingly resemble the input images (i.e., their shadow) and conversely if participants slowed their movements or remained still the outputs would become increasingly influenced by the AI.

**Dynamic Text Prompts**

We began to experiment with altering text prompts in real-time, both programmatically and controlled by an operator, For the installation at *Electromagnetic Field* a large 'playlist' of prompts was used while for the *Storytellers + Machines* installation we developed a system for managing prompt 'scenes'. The latter system implemented a 'base prompt' which defined a style or aesthetic of the scene and then between 3 and 20 additional prompts that fade in and out, with their effects on the output image blending into each other as the scene progressed. Scene progression was tied to participant movement, allowing participants to explore a single moment in the scene by moving slowly, or progress the scene by moving rapidly.

This approach allowed participants to experience the fluid transitions between prompts and create moments of surprise and visual narration while also revealing how the possibility of a given scene was structured. This resonates with findings from Shadowplay V1, however in V2 it was realised at a much quicker pace, experienced as a stream of video rather than as single frames with significant pauses between updates.

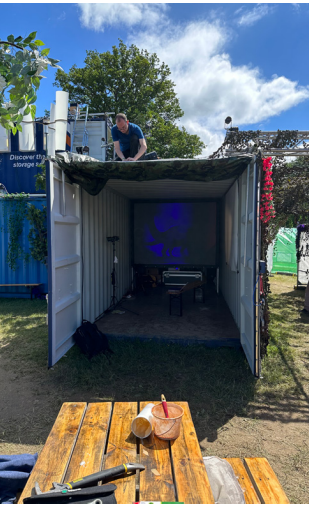
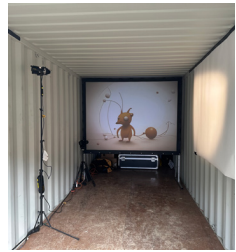
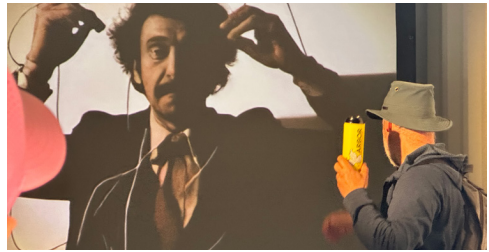
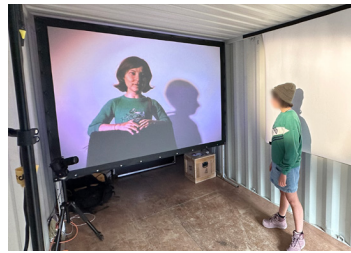
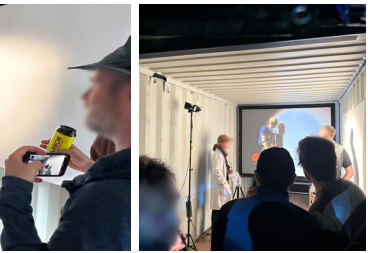

Electromagnetic Field 2024

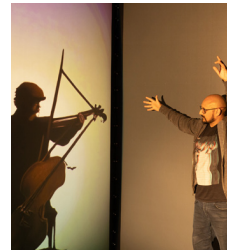
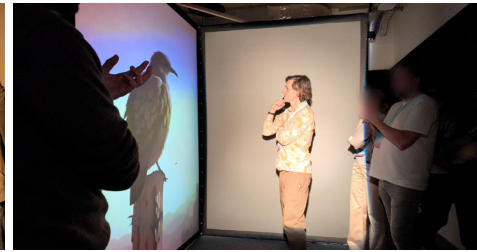
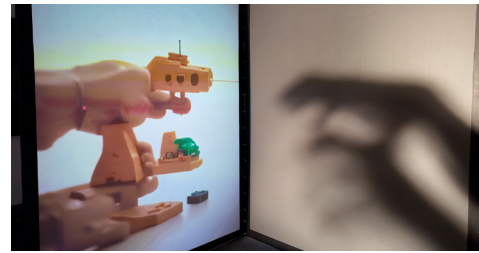
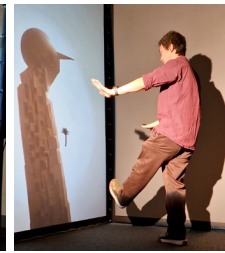
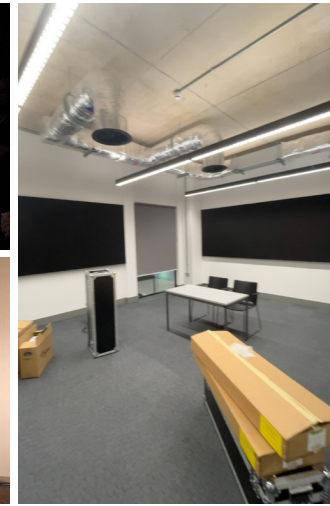

Storytellers + Machines 2024

**Reactive Audio**

We implemented two ways to incorporate sound that is responsive to how participants interact with the system. One of these systems modulates the speed and filtering of a prerecorded music playlist and the other controls the progression of a reactive music system. These implementations of dynamic features were made feasible and desirable by the drastically increased frame rate and the effect of real-time light prompting on the experience. Both systems are driven by participant movement; faster movements increase the pacing of the music and vice versa, producing a bespoke sonic experience accentuating that movement impacts the overall outputs of Shadowplay.

**Holistic Control of the Possibility Space**

The possibility space of Shadowplay V2 emerges from the interplay between a participant's movement, the light prompting (i.e., what image is fed via a camera to the image-to-image AI processor), and how text prompts are implemented. In many cases, these aspects are interrelated (e.g., a participant's movement affects light prompting, changes the AI's parameters, and changes the text prompts). We noted in testing that slight changes in any of these parameters can have significant feedback effects, resulting in drastically different types of output. To manage this situation, we needed to design and develop a user interface. The interface allows fine control over the light prompting, how light prompting drives other parameters, and a means to manually control and automate dynamic text prompts. As the UI developed, we exposed more of the underlying networks and code through the UI, and in turn found we were using them to shape a dynamic image and text-based prompt space.

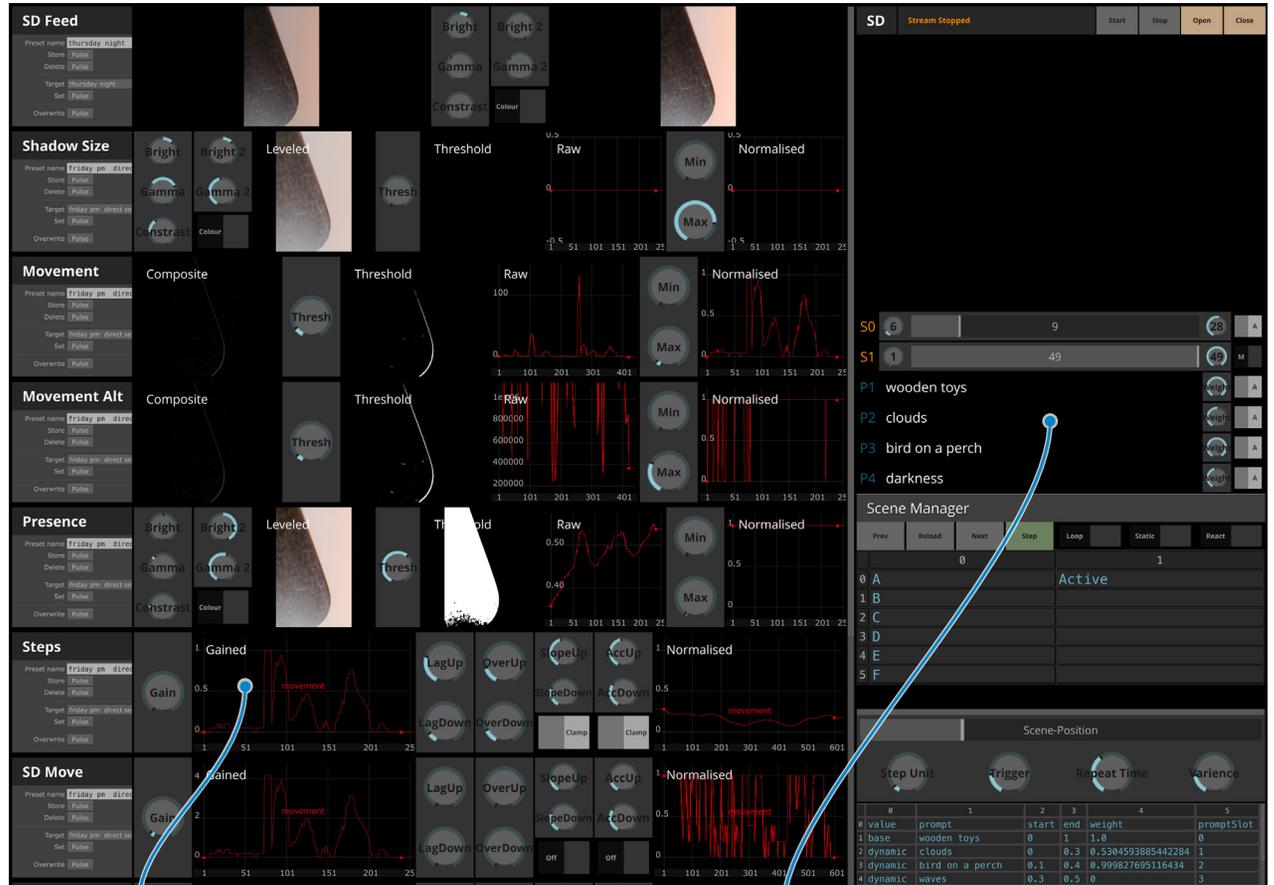

*Signal processing chains*

The left side of the UI c ontrols for various signal chains within the system, starting (at the top) with exposure control over the input video camera signal. Other signal chains shown here quantify the size and movement of shadows across time, and provide controls to smooth and modulate these parameters.

The signals are then fed into StreamDiffusion to control the behavior of the model and used to drive other reactive parts of the system including prompt control, music and audio effects.

*Stream and prompt control*

The right side of the UI provides a real-time display of prompts being fed to StreamDiffusion. Toggle controls allow the user to take manual control over these variables (and enter text-based prompts on the fly), or give control over prompts and weightings to an automated system that stores Prompt Scenes.

The UI allows the user to place the system in a fully automatic mode, which cycles between Scenes automatically as people interact with the piece.

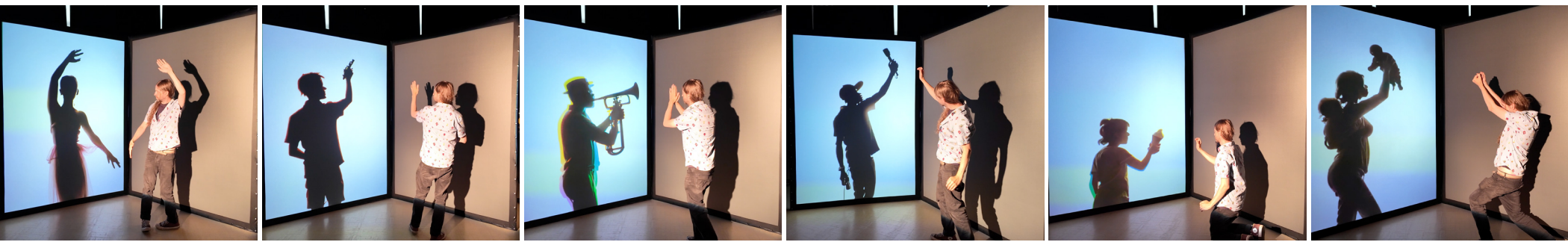

Progression through a prompt scene

**Contrasting the Shadowplay Installations**

Our experimental implementations of dynamic elements in Shadowplay V2 highlighted that as the number of variables and elements that dynamically control how the AI creates the output images increases, the complexity exponentially increases. These complexities were evident at the *Electromagnetic Field* installation, where at times it was difficult to understand how light, camera, UI, and AI model were interacting—this was exacerbated by the context, which included changing lighting conditions throughout the day, multiple people interacting at the same time, and participants positioning themselves directly in front of the camera (as was the case in Shadowplay V1) rather than allowing the camera to capture only shadows.

For the installation at *Storytellers + Machines*, which was in a much more controlled gallery environment as compared to in a steel container at an outdoor festival, we chose to control many of these factors. We achieved this by moving the camera behind a screen, meaning that only shadows could be captured, and people no longer had the option to position themselves in front of the camera. This meant taking control of how light prompting affected the system and in turn more control over possibility space. This aimed to increase the overall legibility of the experience, and better articulate the participants that it was their shadow that was driving the output. This installation also introduced fully reactive music which utilised sound as a cue to encourage awareness of the speed of interactions, which had the consequence of highlighting how movement controls the diffusion amount and therefore how much AI transformation is present in the output image.

Both installations engaged the audience and were enjoyable, with the former being more chaotic and encouraging unadulterated play and the latter encouraging a more introspective and deliberately expressive kind of interaction. It is clear that by affording many dimensions of dynamism the increased frame rate of StreamDiffusion makes increasingly complex tangible and embodied interactions with generative AI a possibility.

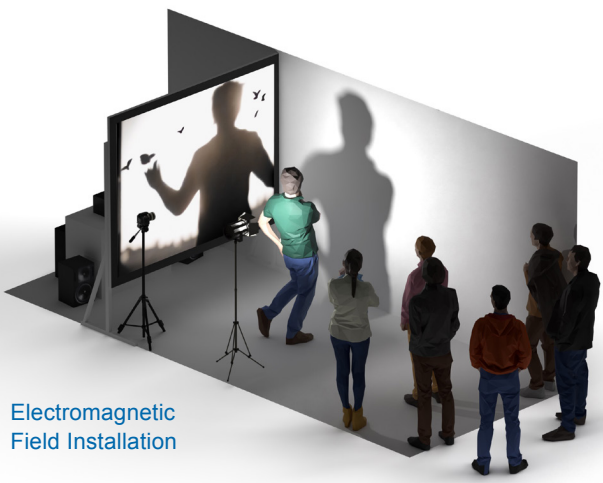

Electromagnetic Field Installation

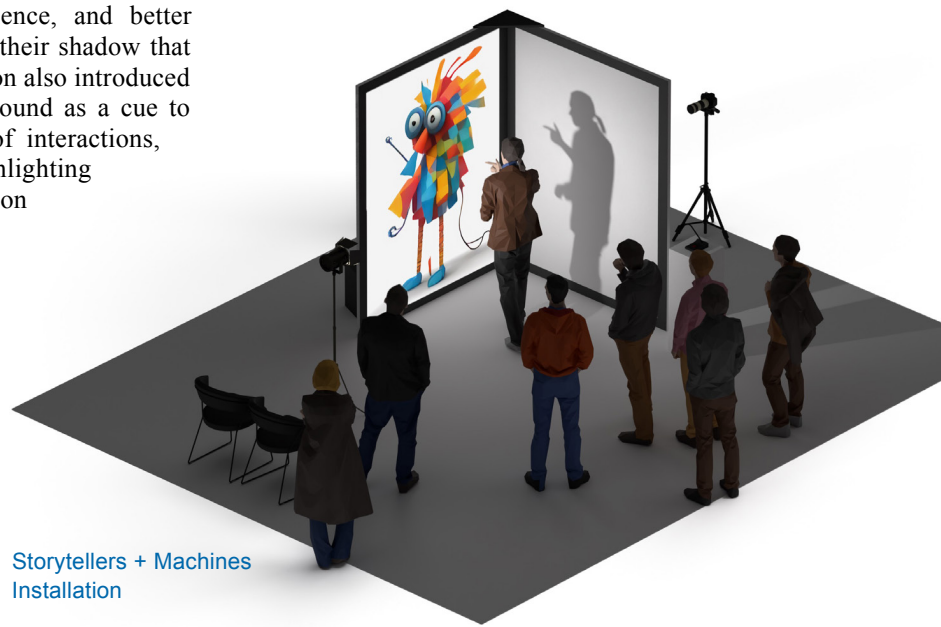

Storytellers + Machines Installation

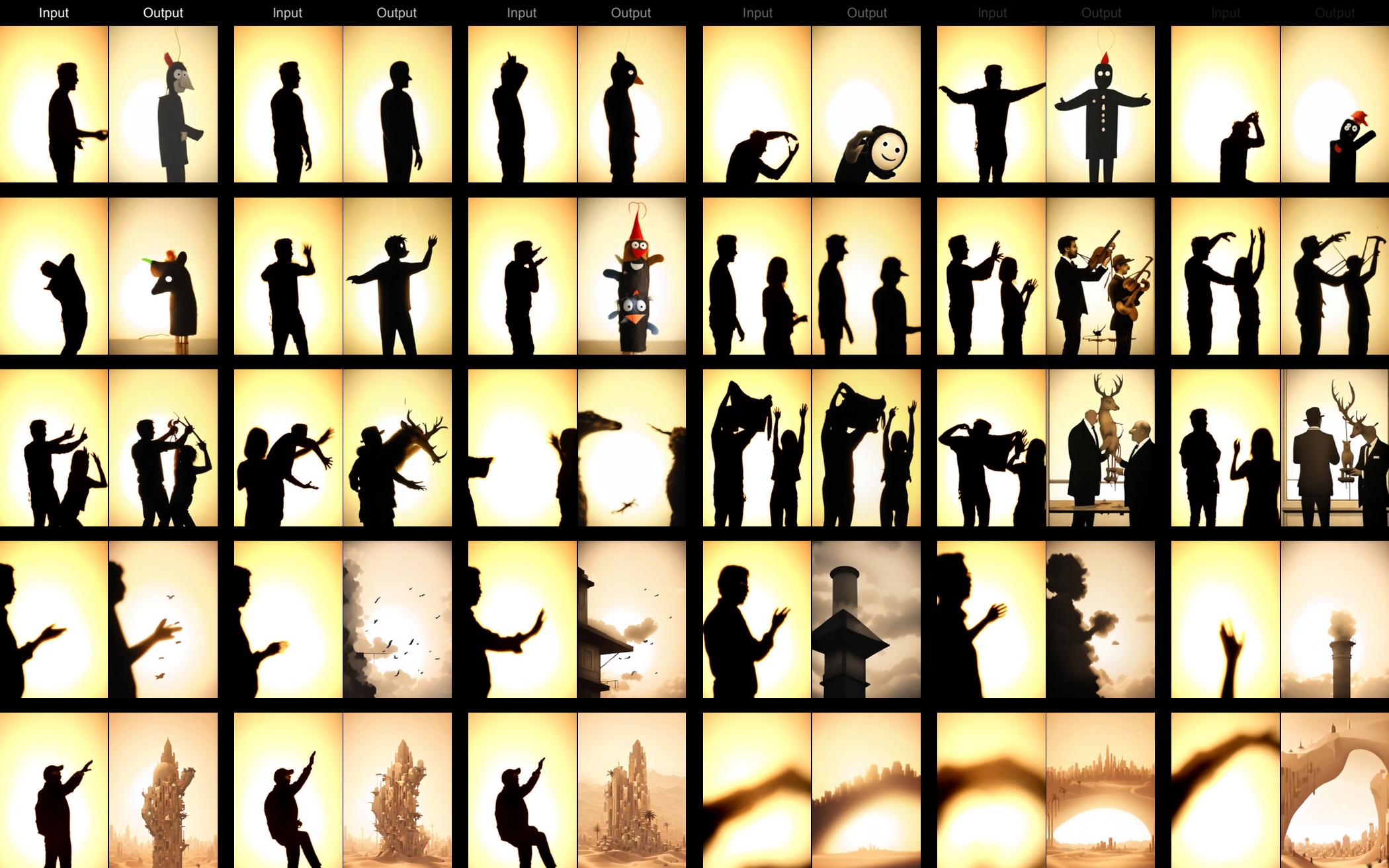

Images from Shadowplay V2 visualising the relationship between input and output Each row represents a span of 1 minute in time

## DISCUSSION

This pictorial has presented three projects that form part of a Research through Design programme exploring the affordances of diffusion-based AI image generation systems. Our programme sits alongside thousands of scholars and practitioners world-wide who are using generative AI across a plethora of contexts and collectively are learning the boundaries and possibilities of this new class of technology.

We presented Shadowplay V1, introducing the concept of light prompting, the idea of using light and shadows to create input images whose attributes provide increased control over the latent space in the model. This was a useful strategy to smooth the embodied interactions Shadowplay facilitates, improving the overall user experience. This technique allowed us to constrain the potential of the AI model's latent space, resulting in a more refined exhibition experience and a consistent aesthetic. Shadowplay V1 also demonstrated how a physical/embodied interaction modality could be used to navigate and explore the vast possibility space of generative AI models.

The second project we presented, Cardshark, allowed us to discuss an iterative and reflexive co-evolution of model choice, prompt fragment selection, and meta-prompt development. This process resembled a craft-like exploration of the AI's material properties. A similar process took place across all the projects, but in developing Cardshark it was particularly stark and allowed us to frame a proposed light-touch method which balances systematic exploration and creative experimentation for some aspects of development and design with generative AI systems.

Shadowplay V2 is defined and characterised by a drastic increase in performance, operating at speeds up to 36x faster than Shadowplay V1 and over 10x faster than Cardshark. To produce a engaging exhibition experience, we had to develop strategies to harness and utilise the dynamism made possible by this increase in performance. Specifically, the project highlighted the need for more sophisticated ways to control the latent possibility space of the AI model while opening a cornucopia of possibilities for variations on tangible and embodied interaction of aspects of the AI. The implementation of dynamic elements such as real-time control of AI parameters, dynamically changing textual prompts, and reactive audio demonstrated the potential for creating rich, responsive interactions with and for generative AI.

Reflecting on these projects we proffer two key contributions:

First, our work provides a perspective on the notion of materiality with respect to diffusion-based generative AI and proposes a method for a craft-like navigation of the latent space within generative AI models. Metaphorically casting data, software, and intangible elements of computing as material is not a new idea. However, the difficult-to-comprehend scope of latent possibility space within generative AI models gives the material metaphor a new relevance. The Shadowplay installations made these qualities evident to both creators and audience members. The installations demonstrated 'soft edges' to the model that could be manipulated, played with, and worked with experientially. Alongside the embodied interactions facilitated an entirely new way to 'scratch the surface' of the latent possibility space, revealing layers in a way that other interactions would not have been able to do so. Furthermore, we propose that the underlying materiality of the AI model was the reason why our findings from the Cardshark project showed that adopting reflective practices of craft, combined with more traditional design approaches, is a valuable strategy when incorporating generative AI into designed activities and products.

Secondly, we provide some additional commentary on interaction design strategies for designing user interfaces informed by the affordances of generative AI. While traditional interaction design methods largely transfer to contexts involving generative AI, the vast possibility space of large AI models necessitates careful control and constraint to create engaging experiences for participants new to the technology. Strategies like light prompting can be used to determine aesthetics, styles, and offer a result similar to the quantization that takes place during model development, but locally and based on the specific requirements of the context. We suggest that creating developmental interfaces that move beyond the current state-of-the-art for generative AI (which typically consists of prompt text boxes, sliders, and buttons) may be a useful strategy to develop systems that are not just usable but help users to arrive at novel outcomes that they otherwise would not encounter. By allowing the parameters of the model—and therefore the latent possibility space—to be controlled through embodied interactions, unexpected parts of the model can be exposed, enabling productive relationships to be surfaced, understood, and meaningfully leveraged. Relatedly, the embodied interactions of Shadowplay and the tangible interactions of Cardshark demonstrate that it is possible to make interactions with generative AI inclusive for a wide range of people. Taken as intermediate knowledge, we hope that this will inspire interaction designers and scholars to produce the next generation of interfaces for generative AI systems, making them more accessible and engaging for diverse user groups.

## FUTURE RESEARCH

When considering directions for future research in this area, the discussion must be tempered by the realisation that innovation is rapid and new possibilities are constantly becoming available. An example of this is the 36x performance increase in the technology during the less-than-12-month span of these projects. Notwithstanding the flux state of the field due to rapid innovation, we do have some immediate next steps in mind for Shadowplay and Cardshark. In both instances, we are exploring new ways of interacting with the underlying prompts that shape the outputs. While the prompt fragments used in Cardshark were successful, the results were very much standalone images. In future research, we will explore how to facilitate the creation of more fully-formed visual stories, which may comprise of changing or multiple images. In the case of Shadowplay we are developing new ways to enable the craft-like prompt development that we established during Cardshark in such a way that prompts can be developed seamlessly while simultaneously carrying out an embodied interaction. In both instances, we are exploring the use of a voice control modalities.

**PROMPT CRAFTING IN PRACTICE**

While it is tempting to offer specific steps and guidance for how to put this kind of thinking into practice, in line with our claim that our contributions are strong concepts or intermediate knowledge, our main intention is to position the ideas in this pictorial as inspirational and generative. Rather than attempting to define precisely how, we hope to inspire other designers, researchers and scholars to adopt a craft approach to prompting. While we do offer some methodological insights relating to Cardshark, it is clear that prompt crafting is likely to always be a negotiation between the context, the specific technologies and models being used, and the learning that takes place as part of the crafting process. Shadowplay is an evolving project with the current codebase and ongoing updates being available to all via https://github.com/designresearchworks/shadowplay

**CONCLUSION**

This design-led research programme seeks to explore and refine approaches for working with generative AI in interactive contexts and report them as a form of intermediate knowledge. By framing interactions with generative as prompt craft rather than prompt engineering, we emphasise the creative, interactive, and exploratory processes that are necessary aspects of developing with generative AI. These attributes are afforded by the underlying materiality of generative AI systems. As the field of generative AI continues to evolve rapidly, we believe that design-led approaches, as well as tangible and embodied interactions, will play a crucial role in shaping how these technologies are integrated into interactive experiences and creative practices.

**ACKNOWLEDGEMENTS**

This work was supported by UK Research and Innovation (UKRI) grant with reference number MR/T019220/1, this Future Leaders Fellowship grant funds the Design Research Works project (https://designresearch.works/).

We would also like to acknowledge the many participants in this project who helped shape the craft we describe; exhibition attendees, workshop participants and especially LICA technical staff at Lancaster University.